\documentclass[12pt]{article}
\usepackage{epsfig}
\input epsf.sty
\input axodraw.sty
\newcommand{\Tint}[1]{{\hbox{$\sum$}\!\!\!\!\!\!\int}_{\!\!\!\!#1}}

\newcommand{\la}[1]{\label{#1}}
 
\newcommand{\be}{\begin{equation}}
\newcommand{\ee}{\end{equation}}
\newcommand{\ba}{\begin{eqnarray}}
\newcommand{\ea}{\end{eqnarray}}
\newcommand{\bi}{\begin{itemize}}
\newcommand{\ei}{\end{itemize}}
\newcommand{\rmi}[1]{{\mbox{\scriptsize #1}}}
\newcommand{\nr}[1]{(\ref{#1})}
\newcommand{\tr}{{\rm Tr\,}}

\newcommand{\im}{\mathop{\rm Im}}
\newcommand{\nn}{\nonumber \\}
\newcommand{\fr}[2]{{\frac{#1}{#2}}}

\renewcommand{\vec}[1]{{\bf #1}}

\newcommand{\RR}{{\rm I\kern -.2em  R}} 
\newcommand{\eq}{Eq.~}
\newcommand{\eqs}{Eqs.~}
\newcommand{\fig}{Fig.~}

\newcommand{\se}{Sec.~}

\def\lsi{\raise0.3ex\hbox{$<$\kern-0.75em\raise-1.1ex\hbox{$\sim$}}}
\def\gsi{\raise0.3ex\hbox{$>$\kern-0.75em\raise-1.1ex\hbox{$\sim$}}}
\newcommand{\lsim}{\mathop{\lsi}}

\makeatletter
\renewcommand\section{\@startsection {section}{1}{\z@}%
                                   {-5.5ex \@plus -1ex \@minus -.2ex}
                                   {2.3ex \@plus.2ex}%
                                   {\normalfont\large\bfseries}}
\renewcommand\subsection{\@startsection{subsection}{2}{\z@}%
                                     {-3.25ex\@plus -1ex \@minus -.2ex}%
                                     {1.5ex \@plus .2ex}%
                                     {\normalfont\normalsize\bfseries}}
\renewcommand\thesection {\@arabic\c@section}
\renewcommand\thesubsection   {\thesection.\@arabic\c@subsection}
\renewcommand{\@seccntformat}[1]{%
\csname the#1\endcsname.\hspace{1.0em}}
\makeatother

\begin{document}
 
\begin{titlepage}
\begin{flushright}
BNL-NT-01/17\\
CERN-TH/2001-164\\
hep-ph/0108034\\
\end{flushright}
\begin{centering}
\vfill
 
\mbox{\bf FINITE BARYON DENSITY EFFECTS ON GAUGE FIELD DYNAMICS} 

\vspace{0.8cm}

D. B\"odeker$^{\rm a,}$\footnote{bodeker@bnl.gov},  
M. Laine$^{\rm b,}$\footnote{mikko.laine@cern.ch} 

\vspace{0.3cm}
{\em $^{\rm a}$%
Department of Physics and RIKEN BNL Research Center,\\ 
BNL, Upton, New York 11973,
USA\\}
\vspace{0.3cm}
{\em $^{\rm b}$%
Theory Division, CERN, CH-1211 Geneva 23,
Switzerland\\}

\vspace*{0.8cm}
 
\end{centering}
 
\noindent
We discuss the effective action for QCD gauge fields 
at finite temperatures and densities, obtained after
integrating out the hardest momentum scales from the system. 
We show that a non-vanishing baryon
density induces a charge conjugation (C) odd operator to the 
gauge field action, proportional to the chemical potential. Even though 
it is parametrically smaller than the leading C even operator, it could 
have an important effect on C odd observables. The same operator appears 
to be produced by classical kinetic theory, allowing in principle 
for a non-perturbative study of such processes.
\vfill
\noindent
 

\vspace*{1cm}
 
\noindent
BNL-NT-01/17\\
CERN-TH/2001-164\\
August 2001 

\vfill

\end{titlepage}


\section{Introduction}

The properties of QCD at a finite temperature $T$ and chemical 
potential $\mu$ play an important role for the physics of the Early 
Universe, heavy ion collision experiments, and neutron stars. 
Despite the presence of a potentially large scale, 
say $T\gg \Lambda_\rmi{QCD}$, QCD however remains
sensitive to non-perturbative 
infrared physics under these conditions~\cite{linde,gpy}. 
This means that
various effective theories may be derived systematically, but the rich 
phenomena encoded in them (for a recent review, see~\cite{rw}) 
may have to be addressed non-perturbatively. 

Quite generically, effective theories tend to possess extra symmetries, 
only broken by higher dimensional operators. A familiar example is weak 
interaction induced strangeness violation in zero temperature QCD. 

Similar phenomena take place also in high temperature physics. 
For instance, the dimensionally reduced effective field theory~\cite{dr} 
describing the thermodynamics of the electroweak sector of the Standard Model 
has extra symmetries: parity (P) and charge conjugation (C) are 
only broken by higher dimensional operators~\cite{parity}. 

In QCD, perturbative interactions break neither P nor C, 
but the latter can be broken by the thermal ensemble, 
if there is a finite baryon density ($\mu\neq 0$). 
Indeed, in the dimensionally reduced effective field theory for the 
thermodynamics of QCD, there is again a new C odd operator, 
of higher order than the leading C even operator~\cite{mu}.

The purpose of this paper is to address the same phenomenon 
in the context of the so called Hard Thermal Loop effective theory
(for a review and references, see~\cite{rob}), the generalisation of 
the dimensionally reduced theory for real time observables. 
We show that a purely gluonic C odd operator is induced 
by quark loops. Thus C odd observables may be directly 
sensitive to non-perturbative bosonic dynamics.  

The environment in which we may envisage our results to 
have significance is mainly that of heavy ion collision 
experiments. In cosmology the baryon chemical potential is 
too small to have any significance, $\mu/T \sim 10^{-8}$. 
In neutron stars, on the other hand, the interesting part of
non-perturbative QCD dynamics 
appears to be 
related more to quark pairing near the Fermi 
surface than to gluons~\cite{rw}. 

As far as heavy ion collisions go, 
precision studies of dimensional reduction show
that effective theories of the type considered may be
quantitatively accurate 
down to $T \sim 2 T_c$~\cite{mu,adjoint,fop,hp,rnew}. At the same 
time, the infrared dynamics described by the effective theories 
is completely non-perturbative and does not allow 
for a perturbative treatment at 
any reasonable temperature~\cite{linde,gpy,mu,adjoint,hp,bn,mdebye,lp}. 
(Apart perhaps from observables such as 
the free energy density where non-perturbative effects 
happen to almost 
cancel out numerically~\cite{a0cond}.) Direct four-dimensional (4d) 
lattice simulations
are not available either for real time quantities
(for a review, see~\cite{db}). Thus effective
theories appear presently to be the only way of studying
quantitatively some interesting 
non-perturbative processes for phenomenologically relevant 
temperatures. 

Among the C odd processes one could think of are that various
correlation lengths, decay rates and oscillation frequencies change, 
because previously distinct quantum number channels 
couple to each other in the presence of $\mu\neq 0$~\cite{ay}. 
In particular the fluctuations of baryon and 
energy densities are correlated. One could also consider 
direct C odd observables, such as 
$M = \sum_{\pi^\pm} 
({\vec{p}_+^2 - \vec{p}_-^2})/({\vec{p}_+^2 + \vec{p}_-^2})$~\cite{bka,kpt}, 
where the sum is over the momenta of all charged pions. 
In addition one may consider 
contributions from C odd amplitudes to C even observables
such as dilepton production~\cite{mg}. 
We do not here study any of these applications, though. 

The plan of the paper is the following. We fix our basic 
notation and conventions in~\se\ref{se:conv}, review briefly 
the effective theory in the static limit in~\se\ref{se:static}, 
and present our derivation of the general non-static case 
in~\se\ref{se:nonstatic}. In~\se\ref{se:clas}
we argue that the result of~\se\ref{se:nonstatic} is equivalent
to another effective description, ``classical kinetic theory'', 
which also allows for a simple numerical lattice implementation. 
We conclude in~\se\ref{se:concl}.

\section{Conventions}
\la{se:conv}

The partition function of QCD at a finite temperature $T$ and chemical 
potential $\mu$ is 
\ba
Z = \tr e^{-\beta (H - \mu Q)} & = & 
\int {\cal D} A_\mu {\cal D}\psi {\cal D}\bar\psi
\exp\Bigl(-\frac{1}{\hbar} S_E  \Bigr), \la{e1} \\
S_E & \equiv & \int_0^{\beta\hbar}\! d\tau \int\! d^3 x\, {\cal L}_E, 
\la{e3} \\
{\cal L}_E & = & 
\fr12 \tr F_{\mu\nu} F_{\mu\nu} + 
\bar\psi [\gamma_\mu D_\mu - \gamma_0 \mu]\psi, \la{action} 
\ea
where $\beta = T^{-1}$, $H$ is the Hamiltonian, $Q$ is the quark number 
operator (three times the baryon number operator),  
boundary conditions over the time direction are periodic for 
$A_\mu$ and antiperiodic for $\psi, \bar\psi$, $A_\mu = A_\mu^a T^a$, 
$\tr T^aT^b = \delta^{ab}/2$,
$D_\mu = \partial_\mu - i g A_\mu$,  
$F_{\mu\nu} = (i/g)[D_\mu,D_\nu]$, and the metric is $(++++)$. 
We set $\hbar = 1$ in the following. As we see from \eq\nr{action}, 
the chemical potential corresponds in momentum space to shifting
fermionic Matsubara frequencies $\omega_n^{f}$ 
as $\omega_n^{f} \to \omega_n^{f} + i\mu$.

The observables we are fundamentally interested in are 
real time correlation functions of the type
\be
C(t_i,x_i;T,\mu) = Z^{-1} \, \tr e^{-\beta (H - \mu Q)}
\Bigl[ O_1(t_1,x_1) O_2(t_2,x_2) ... \Bigr], 
\ee
where $O_i$ are operators in the Heisenberg picture.
As usual, such expectation values can be found by computing first the
corresponding objects in the Euclidean theory of \eqs\nr{e1}--\nr{action},
and performing then an appropriate analytic continuation.

In view of the analytic continuation, we will 
at a number of points already write the action 
in Minkowski notation corresponding to 
the metric $g^{\mu\nu} = {\rm diag}\,(+---)$. 
The continuation will be understood to 
be made by writing $\tau = i t, \partial_\tau = -i \partial_t$, 
$\omega_n = -i \omega$, $A_0^{E} = -i A_0^{M}$. An additional 
minus sign is inserted in the relation 
${\cal L}_E(\tau = i t) = -{\cal L}_M$, such that 
\be
-S_E  = -\int_0^{\beta\hbar} \! d\tau d^3x\, {\cal L}_E\to
i \int dt d^3x {\cal L}_M \equiv i S_M. \la{MinEc}
\ee
In the continuation scalar products change as 
\be
a_E\cdot b_E = a_\mu b_\mu \to - a_M \cdot b_M = - a_\mu b^\mu,
\ee
where we use the implicit notation that both indices
down implies Euclidean metric. 

We denote the various integrations arising as follows:
\ba
\int_x & = & \int d^3x\,dt \; ; \\
\int_\vec{p} & = & \int \frac{d^3p}{(2\pi)^3}\; ; \\ 
\int_P & = & \int_\vec{p} \int \frac{dp_0}{(2\pi)}, 
\quad P_\mu = (p_0, \vec{p})\; ; \\
\Tint{P} & = & \int_\vec{p}\; 
T\sum_{\omega_n} , \quad P_\mu = (\omega_n,\vec{p})\; ,
\ea 
where $\omega_n$ are fermionic ($\omega_n^f$) or bosonic
($\omega_n^b$) Matsubara frequencies; and
\be
\int_v =  \int \frac{d \Omega_v}{4\pi}, \quad v^\mu = (1, v^i), 
\quad v_\mu v^\mu = 0, 
\ee
where the integral is over the directions $\Omega_v$ of $v^i$.
The corresponding $\delta$-functions are assumed 
normalized such that $\int_x \delta_x = 1$. We also denote
\be
\delta_+(P^2) \equiv 2 \theta(p_0)\, (2\pi)\, \delta(P^2), 
\quad P^2 \equiv P\cdot P,  
\la{deltap}
\ee
such that 
\be
\int_P \delta_+(P^2)\, f(p_0,\vec{p}) = 
\int_\vec{p} \frac{f(\omega_P,\vec{p})}{\omega_P}, 
\quad \omega_P\equiv |\vec{p}|. \la{eq:plusint}
\ee

All our results will factorise in a form where a phase space integral 
is left over, which can be carried out explicitly. Let 
\be
n_\rmi{F} (\omega_P) = \frac{1}{e^{\beta \omega_P} + 1}, \quad
N_\pm(\omega_P) = n_\rmi{F} (\omega_P-\mu)\pm n_\rmi{F} 
(\omega_P+\mu). \la{eq:n}
\ee
Then, irrespective of the relative magnitudes of $T,\mu$,
\ba
\int_\vec{p} N_-(\omega_P)  \!\! & = & \!\! 
\frac{\mu}{6}\Bigl(T^2 + \frac{\mu^2}{\pi^2} \Bigr), \la{int1} \\ 
\int_\vec{p} \frac{N_+(\omega_P)}{\omega_P}   \!\! & = & \!\! 
-\fr12
\int_\vec{p} \frac{\partial N_+(\omega_P)}{\partial \omega_P} 
 =  \fr12 \frac{\partial}{\partial\mu}
\int_\vec{p} N_-(\omega_P) = 
\frac{1}{4}\Bigl(\frac{T^2}{3} + \frac{\mu^2}{\pi^2} \Bigr), \la{int2} \\ 
\int_\vec{p} \frac{N_-(\omega_P)}{\omega_P^2}
  \!\! & = & \!\! 
- \! \int_\vec{p} \frac{1}{\omega_P} 
\frac{\partial N_-(\omega_P)}{\partial\omega_P} =  
\fr12 \! \int_\vec{p} \frac{\partial^2 N_-(\omega_P)}{\partial\omega_P^2}
 = \fr12 \! \frac{\partial^2}{\partial\mu^2}
\int_\vec{p} N_-(\omega_P) = \frac{\mu}{2\pi^2}. \la{int3}
\ea

\section{Effective action at the static limit}
\la{se:static}

We start the actual computation by recalling the
fermion induced C odd operators in the static 
limit. This case corresponds to dimensional reduction~\cite{dr}. 

In Minkowski notation, charge conjugation can be defined
for the bosonic fields as 
\be
A_\mu \to -A_\mu^*, \quad
D_\mu \to D_\mu^*, \quad
F_{\mu\nu} \to -F^*_{\mu\nu}.
\ee
Furthermore 
the Minkowski action $S_M$ should be real. Thus one would
expect C odd operators
to have an odd number of gauge fields. In a non-Abelian
case no operator with a single power exists, and hence 
the lowest non-trivial order is cubic. 

Indeed, apart from renormalisation effects, 
the fermionic contribution (with $N_f$ flavours) to the 
effective action of the bosonic Matsubara zero modes is
\be
\frac{\delta {\cal L}_{E}^{f}}{N_f} = 
-i g\frac{\mu}{3} \Bigl(T^2 + \frac{\mu^2}{\pi^2}\Bigr)
\tr A_0 + \frac{g^2}{2} \Bigl(\frac{T^2}{3} + \frac{\mu^2}{\pi^2}\Bigr)
\tr A_0^2 + i \mu \frac{g^3}{3\pi^2} \tr A_0^3
- \frac{g^4}{12\pi^2} \tr A_0^4. \la{static}
\ee
The quadratic and quartic terms here conserve C. 
The linear term breaks C, but is only relevant for an Abelian theory
where $\tr A_0 = A_0$; 
it has been discussed, e.g.,\  in~\cite{ks96}. 
The cubic term, the first C breaking operator in QCD,
has been discussed in~\cite{mu}. 
There are no higher order operators involving $A_0$ only~\cite{mu,cka}.
There are other higher order C breaking operators, though, such as
\be
\delta {\cal L}_{E}^{f} = - i \mu N_f 
 \frac{g^3}{3\pi^2} 
 \frac{7\zeta(3)}{(4\pi T)^2} 
 \biggl(1 + {\cal O}\Bigl(\frac{\mu}{\pi T} \Bigr)^2 \biggr)
 \tr A_0 F_{ij}^2,
\ee
but their contribution is suppressed at least 
by ${\cal O}(\vec{p}^2/(2\pi T)^2)$, where 
$|\vec{p}|\lsim gT$ is the dynamical scale 
within the effective theory. Therefore we ignore them here.

For future reference, we note that if we define 
$j^\nu_a = -(\delta/\delta {A_\nu^a}) S_M$ in the non-Abelian case, 
and $j^\nu = -(\delta/\delta {A_\nu}) S_M$ in the Abelian, then, 
after the analytic continuation discussed around \eq\nr{MinEc}, 
\eq\nr{static} corresponds to
having in the static limit 
\ba
j_\nu^a & = &  - \delta_{\nu 0}\biggl[ N_f \frac{g^2}{2}
\biggl(\frac{T^2}{3} + \frac{\mu^2}{\pi^2} \biggr) A_0^a + 
\mu N_f \frac{g^3}{4\pi^2} d^{abc} A_0^b A_0^c \biggr],  \la{statref} \\
j_\nu & = &  - \delta_{\nu 0}\biggl[\mu N_f \frac{g}{3}
\biggl(T^2 + \frac{\mu^2}{\pi^2} \biggr) +
N_f g^2
\biggl(\frac{T^2}{3} + \frac{\mu^2}{\pi^2} \biggr) A_0+ 
\mu N_f \frac{g^3}{\pi^2} A_0^2 \biggr],  \la{abstatref}
\ea
where we have left out the contributions from the last term 
in~\eq\nr{static}, going beyond the analysis in this paper.

\section{The non-static case}
\la{se:nonstatic}

We now move to the non-static case, to generalise
the effective action in~\eq\nr{static}. 
To determine the leading bosonic C odd operator in QCD, 
we compute the graph in \fig\ref{fig1}.

\subsection{The 3-point function}
\la{se:lo}


\begin{figure}[t]

\begin{center}
\begin{picture}(80,80)(0,0)

\SetWidth{1.5}
\CArc(40,40)(20,0,360)
\Photon(26,54)(12,68){1.5}{3}
\Photon(54,54)(68,68){1.5}{3}
\Photon(40,0)(40,20){1.5}{3}

\Text(5,68)[r]{$A_\mu(Q)$}
\Text(75,68)[l]{$A_\nu(R)$}
\Text(47,0)[l]{$A_\sigma(S)$}

\end{picture}
\end{center}
\caption[a]{The Feynman graph computed in \se\ref{se:lo}.}
\la{fig1}
\end{figure}
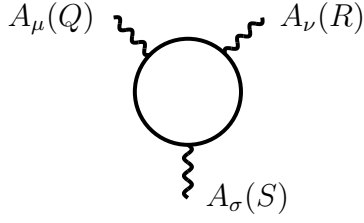


The computation of the graph in~\fig\ref{fig1}
is a straightforward exercise. 
Let us express the result  
as a contribution to the Euclidean action in momentum space.
Then, for an arbitrary gauge field configuration $A_\mu(x)$, we obtain
\ba
\delta S_{E} & = &  \fr43 g^3 N_f \Tint{Q,R,S} \delta_{Q+R+S}\, 
\Gamma_{\mu\nu\sigma}(Q,R,S) \, 
\tr \Bigl[ A_\mu(Q) A_\nu(R) A_\sigma (S) \Bigr]
\la{deltaSE} , 
\ea
where
\ba
\Gamma_{\mu\nu\sigma}(Q,R,S) & = & 
\Tint{P=(\omega_n^{f} + i \mu,\vec{p})}
\frac{F_{\mu\nu\sigma}(p_0,\vec{p})}
{P^2 (P-Q)^2(P+S)^2}, \la{SE3} \\
F_{\mu\nu\sigma}(p_0,\vec{p}) & = &  
\fr14 \tr [\,\slash\!\!\!\! P \gamma_\mu (\slash\!\!\!\! P -\slash\!\!\!\! Q) 
\gamma_\nu (\slash\!\!\!\! P + \slash\!\!\!\! S) \gamma_\sigma ].
\la{gamma} 
\ea
To evaluate the sum over $\omega_n^f$, we use the contour formula
\be
T \sum_{n \rmi{ odd}}f(n \pi T + i \mu) = 
\sum_\rmi{poles at $\im z\neq \mu$} 
\frac{i \mathop{\mbox{Res}} f(z)}{e^{i \beta z + \beta \mu} + 1}. \la{res}
\ee
In addition we denote 
\be
\omega_P^2 = \vec{p}^2, \quad
\omega_{P-Q}^2 = (\vec{p} - \vec{q})^2, \quad
\omega_{P+S}^2 = (\vec{p} + \vec{s})^2.
\ee
To consider only the part odd in C, we replace in the whole expression
\be
f(\mu) \to \fr12\Bigl[f(\mu) - f(-\mu) \Bigr].
\ee 

Picking up the poles according to \eq\nr{res}, and using
momentum conservation as well as the identity
$n_\rmi{F}(-\omega_P) = 1 - n_\rmi{F}(\omega_P)$ (cf.\ \eq\nr{eq:n}),
the expression in \eq\nr{SE3} becomes    
\ba
\Gamma_{\mu\nu\sigma} \!\! & = & \!\! \int_\vec{p} \biggl\{
\frac{N_-(\omega_P)}{4\omega_P}\biggl[
\frac{F_{\mu\nu\sigma}(-i \omega_P,\vec{p})}
{[(q_0+i\omega_P)^2 + \omega_{P-Q}^2][(s_0-i\omega_P)^2 + \omega_{P+S}^2]}
-(\omega_P \to -\omega_P)
\biggr] \\
&  & \hspace{-1cm} {}+
\frac{N_-(\omega_{P-Q})}{4\omega_{P-Q}}\biggl[
\frac{F_{\mu\nu\sigma}(q_0-i \omega_{P-Q},\vec{p})}
{[(q_0-i\omega_{P-Q})^2 + \omega_P^2][(r_0+i\omega_{P-Q})^2 + \omega_{P+S}^2]}
-(\omega_{P-Q} \to -\omega_{P-Q})
\biggr] \nn 
&  & \hspace{-1cm}{}+
\frac{N_-(\omega_{P+S})}{4\omega_{P+S}}\biggl[
\frac{F_{\mu\nu\sigma}(-s_0-i \omega_{P+S},\vec{p})}
{[(s_0+i\omega_{P+S})^2 + \omega_P^2][(r_0-i\omega_{P+S})^2 + \omega_{P-Q}^2]}
-(\omega_{P+S} \to -\omega_{P+S})
\biggr] \biggr\}.
\nonumber 
\ea
In fact each of these terms gives an identical contribution to \eq\nr{deltaSE}:
changing integration variables on the second line such 
that $\vec{p}\to \vec{p}+\vec{q}$ (so that $\omega_{P-Q} \to \omega_P$), 
on the third line such
that $\vec{p}\to \vec{p}-\vec{s}$ (so that $\omega_{P+S} \to \omega_P$), 
and in the terms obtained with 
$(\omega_P\to -\omega_P)$ such that $\vec{p}\to -\vec{p}$, 
we arrive at 
\ba
\Gamma_{\mu\nu\sigma} \!\! & = & \!\! - \int_\vec{p}
\frac{N_-(\omega_P)}{16\omega_P^3}\biggl[
\frac{F_{\mu\nu\sigma}(-i \omega_P,\vec{p})}
{\Bigl(v_E\cdot Q - \frac{Q^2}{2\omega_P} \Bigr)
\Bigl(v_E\cdot S + \frac{S^2}{2\omega_P} \Bigr)}
-
\frac{F_{\mu\nu\sigma}(i \omega_P,-\vec{p})}
{\Bigl(v_E\cdot Q + \frac{Q^2}{2\omega_P} \Bigr)
\Bigl(v_E\cdot S - \frac{S^2}{2\omega_P} \Bigr)} \nn
&  & \hspace*{4cm} + \Bigl(  
\mbox{cyclic permutations of } \mu,\nu,\sigma; Q,R,S\Bigr) 
\biggr], \la{s2}
\ea
where we have denoted $v_{E,\mu} = (-i,p_i/\omega_P)$. But the cyclic
permutations are automatically reproduced by the trace in \eq\nr{deltaSE}, 
so we can replace them by a factor~3.

We now go to Minkowski space as discussed 
in the paragraph around \eq\nr{MinEc}, and 
carry out the small effective coupling expansion. 
In other words, we look for the leading term in the expansion 
in small $Q/\omega_P,R/\omega_P,S/\omega_P$, 
where parametrically $Q,R,S \lsim {\rm max}(gT,g\mu)$, while
the integration variable gets its contributions from 
$\omega_P\sim {\rm max}(T,\mu)$ (cf.\ \eqs\nr{int1}--\nr{int3}). 
Naively, the leading term in \eq\nr{s2} 
is of order ${\cal O}((QS)^{-1})$. This however vanishes after
cyclic permutations, due to momentum conservation
(cf.\ \eq\nr{ide1}). There is 
no term of order ${\cal O}(Q^{-1})$ due to the symmetries of the 
expression, and then the leading term is of order ${\cal O}(Q^0)$.
In order to find it out, we need to know 
$F_{\mu\nu\sigma}$ to order 
${\cal O}(Q^2)$, because it is multiplied 
by a term $\sim {\cal O}((QS)^{-1})$.

Taking traces over Dirac gamma matrices in \eq\nr{gamma}
(already transformed to \linebreak Minkowski space
and with indices raised), we obtain 
\ba
& & \hspace*{-2cm}
F^{\mu\nu\sigma}(-i \omega_P,\vec{p}) \nn
&  = &   
4 \omega_P^3 \Bigl[ v^\mu v^\nu v^\sigma \Bigr] \nn
& + & 2 \omega_P^2 \Bigl[
v\cdot Q\, v^\sigma g^{\mu\nu}
-v\cdot S\, v^\mu g^{\nu\sigma}
-v^\mu v^\sigma Q^\nu-v^\nu v^\sigma Q^\mu 
+v^\mu v^\sigma S^\nu + v^\mu v^\nu S^\sigma \Bigr] \nn
& + & 
\omega_P\Bigl[ 
-v\cdot Q\, (g^{\sigma\nu} S^\mu -
g^{\mu\nu} S^\sigma -g^{\mu\sigma} S^\nu) - 
v\cdot S\, (g^{\mu\nu} Q^\sigma - g^{\nu\sigma}Q^\mu - 
g^{\mu\sigma} Q^\nu) \nn
& & \hspace*{0.6cm} 
-  Q\cdot S\, (v^\nu g^{\mu\sigma} - v^\sigma g^{\mu\nu} -
v^\mu g^{\nu\sigma}) \nn
& & \hspace*{0.6cm} 
- v^\sigma (Q^\mu S^\nu + S^\mu Q^\nu) - 
v^\mu (Q^\sigma S^\nu + Q^\nu S^\sigma) + 
v^\nu (Q^\sigma S^\mu - S^\sigma Q^\mu)
\Bigr], \la{F}
\ea
where now $v^\mu = (1,p^i/\omega_P)$, $v\cdot v = 0$.

The terms odd in $\omega_P$ in \eq\nr{F} pick up the first
and third terms in the expansion of the denominators of \eq\nr{s2}
in $Q/\omega_P,S/\omega_P$, 
while the term even in $\omega_P$ picks up the second term. 
Using in addition momentum conservation to 
simplify the expression, we finally obtain 
\ba
\delta S_M & = & -\fr12 g^3 N_f \int_{Q,R,S} \delta_{Q+R+S} \,
\tr \Bigl[ A_\mu (Q) A_\nu(R) A_\sigma(S) \Bigr] 
\int_\vec{p}
\frac{N_-(\omega_P)}{\omega_P^2}  \nn
& \times & \int_v  \biggl[
\frac{v^\mu v^\nu v^\sigma}{(v\cdot Q)(v\cdot R)}
\biggl(\frac{(Q^2)^2}{(v\cdot Q)^2} - 
\frac{Q^2R^2}{(v\cdot Q)(v\cdot R)}+
\frac{(R^2)^2}{(v\cdot R)^2} \biggr) \nn
& & +  
\frac{v^\mu v^\nu R^\sigma + v^\mu v^\sigma R^\nu
-v^\mu v^\nu Q^\sigma-v^\nu v^\sigma Q^\mu}{(v\cdot Q)(v\cdot R)}
\biggl(\frac{Q^2}{v\cdot Q} - \frac{R^2}{v\cdot R} \biggr) \nn
& & -  2 \frac{v^\sigma}{v\cdot S}
\biggl( g^{\mu\nu} \frac{S^2}{v\cdot S}+ 
\frac{R^\nu S^\mu}{v\cdot R}+
\frac{Q^\mu S^\nu}{v\cdot Q} \biggr) + 
2 \frac{S^\sigma}{v\cdot S} g^{\mu\nu}  \biggr].
\la{s4}
\ea
This result can still be made more transparent, however.  
We write the gauge field as
\be
A_\mu(Q) = \tilde A_\mu(Q,v) +  
\frac{v^\alpha Q_\mu}{v\cdot Q} A_\alpha(Q), \quad
\tilde A_\mu (Q,v) \equiv \biggl({\delta^\alpha}_\mu - 
\frac{v^\alpha Q_\mu}{v\cdot Q} \biggr) A_\alpha(Q), \la{Atilde} 
\ee
such that $v\cdot \tilde A = 0$. It is then 
straightforward, employing the
momentum conservation identity 
\be
\frac{1}{(v\cdot    Q)(v\cdot  R)}+
\frac{1}{(v\cdot    Q)(v\cdot  S)}+
\frac{1}{(v\cdot  S)(v\cdot  R)} = 0, \la{ide1} 
\ee
to show that only terms involving $\tilde A_\mu(Q,v)$
survive in \eq\nr{s4}. Furthermore, since  $\tilde A_\mu(Q,v)$
is transverse with respect to $v^\mu$, only the last term 
in \eq\nr{s4} gives a contribution. This result is what would
have been obtained if we had relied on the 
gauge choice $v\cdot \tilde A = 0 $~\cite{fgt,elmfors} to begin with. 
Going furthermore to $x$-space, we arrive at
\ba
\delta S_M & = & - g^3 N_f \int_\vec{p} \frac{N_-(\omega_P)}{\omega_P^2}
\int_{x,v} 
\tr \Bigl[ \tilde A_\mu \tilde A^\mu
\Bigl(\frac{1}{v\cdot \partial} \partial^\sigma \tilde A_\sigma \Bigr)\Bigr]  
\nonumber  \\
 & = & -\frac{1}{2\pi^2} g^3\mu N_f 
 \int_{x,v} 
\tr \Bigl[ \tilde A_\mu \tilde A^\mu
\Bigl(\frac{1}{v\cdot \partial} \partial^\sigma \tilde A_\sigma \Bigr)\Bigr],
\la{s5}  
\ea
where the last integral was from~\eq\nr{int3}.
This is the C odd part of~\fig\ref{fig1} 
at leading order in the small coupling expansion,
for  $Q,R,S\lsim  {\rm max}\, (gT,g\mu) \ll {\rm max}\, (T,\mu)$. 

\subsection{Higher point functions}
\la{se:ho}


\begin{figure}[t]

\begin{center}
\begin{picture}(210,40)(0,0)

\SetWidth{1.5}
\Line(0,0)(210,0)
\Photon(30,0)(30,20){1.5}{3}
\Photon(80,0)(80,20){1.5}{3}
\Photon(130,0)(130,20){1.5}{3}

\Text(30,28)[b]{$A_\mu(Q)$}
\Text(80,28)[b]{$A_\nu(R)$}
\Text(130,28)[b]{$A_\sigma(S)$}
\Text(180,13)[b]{$\cdots\cdots$}
\Text(-5,0)[r]{$\vec{p}$}

\end{picture}

\end{center}
\caption[a]{The Feynman graphs discussed in \se\ref{se:ho}.}
\la{fig2}
\end{figure}
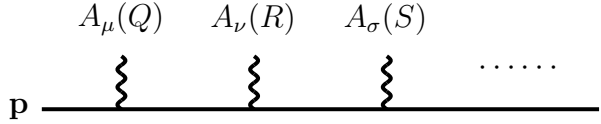


In a non-Abelian theory, 
the effective action in \eq\nr{s5} is not
explicitly gauge invariant. 
To obtain a gauge
invariant result one has to account
also for $n$-point functions with $n > 3$
at leading order in $g$. They contribute to the effective action at
the same order as~\eq\nr{s5}, if we consider non-perturbative
gauge field configurations such that 
$\partial \sim g \tilde A \sim g^2T$. 
Fortunately such higher point functions can easily be computed, 
if we rely on the gauge choice $v\cdot \tilde A = 0$~\cite{fgt,elmfors},
as we now show.

Let us first recall  the idea behind the gauge
choice $v\cdot \tilde A = 0$. 
Once we have carried out the frequency sum in~\eq\nr{res}, 
we are essentially evaluating the propagator of an on-shell fermion
in a background gauge field (see \fig\ref{fig2}). 
Because the fermion is on-shell, the result should be gauge choice independent 
for each $\vec{p}$ separately, or equivalently, for each $v^\mu$.
Thus we can choose a gauge separately for each $v^\mu$ such 
that $v\cdot \tilde A = 0$, carry out a simplified analysis in terms 
of $\tilde A_\mu$, and in the end write the result in a gauge 
invariant form, whereby the gauge choice can be removed. 

In this gauge it is easy to see that the only higher point
1-loop graph which can contribute to the term linear in $\mu$
(such as in \eq\nr{s5}) is 
quartic in $\tilde A$. Indeed, after the frequency
sum, the result for an $n$-point function is parametrically
of the form (cf.~\eq\nr{s2})
\be
\delta S_E \sim \Tint{Q}\, \int_\vec{p} \frac{N(\omega_P)}{\omega_P^n} 
\frac{{\rm Tr}[\, \slash\!\!\!\! \tilde A^n(Q) \, \slash\!\!\!\!
	P^m \, \slash\!\!\!\! Q^{n-m}]}{Q^{n-1}}, 
\ee
where $\slash\!\!\!\! \tilde A = \gamma_\mu \tilde A_\mu$, 
$0\le m\le n$, and $Q$ denotes collectively the soft external momenta.  
Taking the trace one obtains scalar products of 
$ \tilde A $, $ P $ and the $ Q $'s.
Now if $m> n/2$, 
there must be at least one product  
$P^2$ or $P\cdot \tilde A$. Both of them vanish: 
the first because the frequency sum puts us on the pole
(cf.\ \eq\nr{s2}), 
and the second because of the gauge choice. Thus the leading
$\vec{p}$-integral corresponds to $m=n/2$, and is 
\be
\delta S_E \sim \Tint{Q}\, \frac{\tilde A^n(Q)}{Q^{{n}/{2}-1}} 
\int_\vec{p} \frac{N(\omega_P)}{\omega_P^{{n}/{2}}}. 
\ee
But this integral can lead to
a linear dimensionality, i.e.\ a term proportional to~$\mu$,
only for $n\le 4$.

We have thus computed the graph for $n=4$,
\ba
\delta S_E & \propto &  g^4 N_f \Tint{Q,R,S,T} \delta_{Q+R+S+T} \,\,
\Tint{P=(\omega_n^{f} + i \mu,\vec{p})} \nn
& & \times 
\frac{\tr [ \,
\slash\!\!\!\! P \,
\slash\!\!\!\! \tilde A(Q) \,
(\slash\!\!\!\! P  - \slash\!\!\!\! Q) \,
\slash\!\!\!\! \tilde A(R) \,
(\slash\!\!\!\! P - \slash\!\!\!\! Q - \slash\!\!\!\! R) \,
\slash\!\!\!\! \tilde A(S) \,
(\slash\!\!\!\! P  + \slash\!\!\!\! T) \, 
\slash\!\!\!\! \tilde A(T)
]}
{P^2 (P-Q)^2(P-Q-R)^2(P+T)^2}. 
	\label{fourpoint}
\ea
Taking again into account 
that $P^2=0$ after the frequency sum (cf.\ \eq\nr{s2}), 
that because of the gauge choice, $\slash\!\!\!\! P$ 
anticommutes with $\slash\!\!\!\! \tilde A$, 
and that we need the highest power of $P$ possible in the numerator, 
the relevant part of the trace is 
\ba
\tr [\,
\slash\!\!\!\! P \, 
\slash\!\!\!\! \tilde A(Q) \,
\slash\!\!\!\! Q \,
\slash\!\!\!\! \tilde A(R) \,
\slash\!\!\!\! P \,
\slash\!\!\!\! \tilde A(S) \,
\slash\!\!\!\! T  \,
\slash\!\!\!\! \tilde A(T)
] 
= 2 (P \cdot Q) (P \cdot T) 
\tr [\,
\slash\!\!\!\! \tilde A(Q) \,
\slash\!\!\!\! \tilde A(R) \,
\slash\!\!\!\! \tilde A(S) \,
\slash\!\!\!\! \tilde A(T) ]. 
\ea
But then the integral remaining is (written now in Minkowski space)
\be
\delta S_M \sim g^4 N_f  
\int_{Q,R,S,T} \delta_{Q+R+S+T}
\int_\vec{p} \frac{N_-(\omega_P)}{\omega_P^2}
\int_v \frac{\tr [
\slash\!\!\!\! \tilde A(Q) \,
\slash\!\!\!\! \tilde A(R) \,
\slash\!\!\!\! \tilde A(S) \,
\slash\!\!\!\! \tilde A(T) ]}{v\cdot (Q+R)}.
\ee
This however vanishes,
as can be seen using momentum conservation and 
cyclic permutations of the trace. We thus find that the
correct result for the term proportional to $\mu$, 
including all 1-loop corrections, 
is already given by \eq\nr{s5}. 

The specific gauge choice made can now be relaxed by writing 
\be
\tilde A_\mu = \frac{1}{v\cdot {\cal D}} v^\alpha F_{\alpha\mu}, \quad
\frac{ 1}{v \cdot \partial } \partial ^ \sigma \tilde A_\sigma 
= \frac{ 1}{v \cdot {\cal D}} 
{\cal D} ^ \sigma \frac{ 1}{v \cdot {\cal D}} v ^ \gamma 
F _{\gamma \sigma} 
\la{AtoF}
\ee
where ${\cal D}$ is the 
covariant derivative in the adjoint representation.
The right hand sides of Eq.~(\ref{AtoF})  are gauge covariant, and taking 
the trace in Eq.~(\ref{s5}) makes their product gauge invariant. Thus in 
an arbitrary gauge we get
\ba
\delta S_M \!\! & = & \!\! - \frac{1}{2\pi^2}g^3\mu N_f 
\int_{x,v} \!\!
\tr 
\Bigl(\frac{1}{v\!\cdot\! {\cal D}} v^\alpha F_{\alpha\mu}\Bigr)
\Bigl(\frac{1}{v\!\cdot\! {\cal D}} v^\beta {F_{\beta}}^{\mu}\Bigr)
\Bigl(\frac{1}{v\!\cdot\! {\cal D}}{\cal D}^\sigma 
      \frac{1}{v\!\cdot\! {\cal D}} v^\gamma F_{\gamma\sigma} 
\Bigr).
\la{s6}
\ea
This is our final result for
the leading bosonic C odd operator, to be added to 
the usual Hard Thermal Loop action,
given in~\cite{bp}. 

We have not carried out explicitly the angular integral $\int_v$ 
in \eq\nr{s6}. Various such integrals are discussed in~\cite{ft}.
The structure of these integrals is rather non-trivial and it is, 
for instance, not at all obvious that \eq\nr{s6} reduces 
to the third term in~\eq\nr{static} in the static limit, without carrying 
out the integration explicitly. We shall return to this issue
in the next section, where we find a much simpler way of 
checking that the correct static limit is reproduced. 

\section{Classical kinetic theory}
\la{se:clas}

In \se\ref{se:nonstatic} we computed the leading bosonic C odd operator
in the Hard Thermal Loop action, \eq\nr{s6}. The form of this operator 
is however not particularly useful for any practical applications. 
Let us therefore show that the same physics is contained in a much 
simpler description, that of classical 
kinetic theory~\cite{uh} (for a review, see~\cite{he}). 
For notational simplicity we shall work within an Abelian 
theory for the moment, and return to the non-Abelian case 
only in~\se\ref{se:nonab}.

\subsection{Setup}

We follow here the pedagogic presentation in Ref.~\cite{rob}
(based on the original work in~\cite{silin,kelly,bft}). 
Let us start with classical electrodynamics, and define
\be
p^\alpha = \frac{d x^\alpha}{d t}, \quad
\frac{d p^\alpha}{d t} = - g {F^{\alpha}}_\beta p^\beta. 
\ee
Then the collisionless Boltzmann equation for hard particles
in a gauge field background becomes
\be
\frac{d f(x,p)}{dt} = p^\alpha \biggl( 
\frac{\partial f}{\partial x^\alpha} + g {F_\alpha}^\beta
\frac{\partial f}{\partial p^\beta} \biggr) = 0. \la{boltz}
\ee
Let us note that we may in general assume that
\be
f(x,p) = \delta_+ (p^2) \hat f(x,p),  \la{defhat}
\ee
where $\delta_+(p^2)$ is defined as in \eq\nr{deltap}, 
since this form is conserved by \eq\nr{boltz}, due to
\be
p^\alpha {F_\alpha}^\beta \frac{\partial}{\partial p^\beta} 
\delta (p^2) = 2
p^\alpha p^\beta F_{\alpha\beta}
\delta' (p^2) = 0. 
\ee
The derivate of $\theta(p_0)$ included in $\delta_+ (p^2)$ 
can be safely ignored as well, since it would contribute only at the 
point $p_0 = \vec{p} =0$, 
and has no effect after integration over $p$.

We formally solve \eq\nr{boltz} in powers of $g F_{\mu\nu}$: 
$f = f_0 + f_1 + f_2 + ...$ .
This leads to the recursion relation 
\be
p\cdot \partial f_{n+1}(x,p) = - g p^\alpha {F_{\alpha\beta}}(x)
\frac{\partial f_n(x,p)}{\partial p_\beta}. \la{iter}
\ee
The zeroth order gives
\be
p\cdot \partial f_0(x,p) = 0.
\ee 
We take 
as a solution a space-time independent function 
depending, in view of \eq\nr{defhat}, 
non-trivially only 
on $p_0$, parametrized by $T,\mu$, and
applying separately to all particle species $i$:
\be
f_0^{(i)} = \delta_+(p^2)\,\hat f_0^{(i)}(p_0; T,\mu_i) = 
\delta_+(p^2)\,n_\rmi{F}(p_0-\mu_i), \la{eq:f0}
\ee
where $n_\rmi{F}(p_0)$ is in \eq\nr{eq:n}. Furthermore, 
antiparticles are always assumed to come with the opposite 
signs of $g$ and $\mu$ than particles. Thus, 
a single Dirac fermion contributes two degrees of freedom
with $+g,+\mu$, two with $-g,-\mu$.

In addition to these equations, we need the definition of the 
current induced by the hard particles:
\be
j^\mu(x) = - \sum_i g_i \int_{P} p^\mu f^{(i)}(x,p). \la{jmu}
\ee
The equations of motion are 
\be
S_M^\rmi{free} = -\int_x \fr14 F_{\mu\nu}F^{\mu\nu}, \quad
\frac{\delta}{\delta A_\mu} S_M^\rmi{free} =  
\partial_\nu F^{\nu\mu} = j^\mu. 
\ee
The expression for $j^\mu$ in terms of the background gauge
field thus implies a non-local effective action 
$S_M  =  S_M^\rmi{free} + \delta S_M$
for the gauge fields only, where 
$\delta S_M$ is to be determined from 
\be
 \frac{\delta}{\delta A_\mu}
 \delta S_M =   - j^\mu.
\la{SMint}
\ee

\subsection{Linear term}

Let us now work out explicit expressions. 
We start by considering $f = f_0$. 
Summing over $N_f$ flavours, each with two degrees of freedom 
with $+g,+\mu$ and two with $-g, -\mu$, we obtain from
\eqs\nr{eq:f0}, \nr{jmu},
\be
j^\mu = - 2 g N_f \int_P \delta_+(p^2)\, p^\mu N_-(p_0) 
 = -  \delta^{\mu 0}\, 
 g N_f \frac{\mu}{3}\Bigl(T^2 + \frac{\mu^2}{\pi^2} \Bigr),
\ee
where we used \eqs\nr{eq:plusint}, \nr{eq:n}, \nr{int1}.
\eq\nr{SMint} is then trivially solved. The result, 
\be
\delta S_M = - \int_x j^0 A_0,   
\ee
reproduces the first term in~\eq\nr{static} after analytic continuation.

\subsection{Quadratic term}

For the next term the considerations are well-known, 
but for completeness we briefly present them here, too. We need 
\be
f_{1} = - g \frac{1}{p\cdot \partial} p^\alpha {F_{\alpha\beta}}
\frac{\partial f_0}{\partial p_\beta}.
\ee
Inserting into \eq\nr{jmu}, 
\be
j^\mu(x) =  \sum_i g_i^2 \int_{P}
\frac{\partial f_0^{(i)}}{\partial p_\beta} 
\frac{p^\mu p^\alpha}{p\cdot \partial} F_{\alpha\beta}(x). \la{qdjmu}
\ee
As described in~\cite{rob}, this 
corresponds to the usual Hard Thermal Loop action~\cite{bp} 
\be
\delta S_M = - g^2 N_f \int_\vec{p} \frac{N_+(\omega_P)}{\omega_P}\int_{x,v}
(v^\mu F_{\mu\rho}) 
\frac{1}{(v\cdot\partial)^2} (v^\nu {F_\nu}^\rho). 
\la{HTL}
\ee

To check this result at the static limit, 
it is actually convenient to start from the form in 
\eq\nr{qdjmu} rather than that in~\eq\nr{HTL}. We shall 
discuss this is detail for the non-Abelian 
case in~\se\ref{se:nonab}, and following that 
argument, it is easy to see that the second term 
in \eq\nr{abstatref} immediately follows from~\eq\nr{qdjmu}.

\subsection{Cubic term}
\la{se:cubic}

We then proceed to the next order in~$g F_{\mu\nu}$. We are not aware 
of previous analyses going beyond~\eq\nr{HTL} in this way.

To proceed, it is useful to 
Fourier transform with respect to $x$: 
\be
f(x,p) = \int_Q e^{i Q\cdot x} f(Q,p), \quad
f(Q,p) = \int_x e^{-i Q\cdot x} f(x,p).
\ee
Solving for $f_2$ from \eq\nr{iter} and then inserting into \eq\nr{jmu}, 
we obtain 
\ba
j^\mu(Q) & = & - \sum_i g_i^3 \int_{R,S}\int_{P} \delta_{Q-R-S} \, 
\frac{p^\mu}{i p \cdot Q} p^\alpha F_{\alpha\beta}(R) 
\frac{\partial}{\partial p_\beta}\biggl[
\frac{1}{i p\cdot S} p^\delta F_{\delta\gamma}(S) 
\frac{\partial f_0^{(i)}}{\partial p_\gamma}
\biggr]  \nn
& = & 
\sum_i g_i^3 \int_{R,S}\int_{P} \delta_{Q-R-S} \,
\frac{\partial f_0^{(i)}}{\partial p_\gamma}
\frac{\partial}{\partial p_\beta}
\biggl(
\frac{p^\mu}{i p \cdot Q}\biggr) 
p^\alpha F_{\alpha\beta}(R) 
\frac{1}{i p\cdot S} p^\delta F_{\delta\gamma}(S) \nn 
& = & 
-\sum_i g_i^3 \int_{R,S}\int_{P} \delta_{Q-R-S}  \, f_0^{(i)}
\frac{\partial}{\partial p_\gamma}\biggl[ 
\frac{\partial}{\partial p_\beta}
\biggl(
\frac{p^\mu}{p \cdot Q}\biggr)(p\cdot R) 
\tilde A_\beta(R) \tilde A_\gamma(S) \biggr], \la{j3mu}
\ea
where we carried out two partial integrations
and wrote (cf.\ \eq\nr{AtoF}) 
\be
p^\alpha {F_{\alpha\beta}}(R)  = 
i (p \cdot R) \tilde A_\beta(R) = 
i (p \cdot R)^2 {\partial \over \partial p^\beta}
\Big( {p^\nu \over p\cdot R}\Big) A_\nu(R). 
\la{def}
\ee
We can now take the partial derivatives
$\partial/\partial p_\gamma, \partial/\partial p_\beta$, using 
\be
\frac{\partial \tilde A_\mu(Q)}{\partial p^\sigma} = 
- \frac{Q_\mu}{p\cdot Q} \tilde A_\sigma (Q), \la{eq:dAt}
\ee 
following from \eq\nr{def} (or \eq\nr{Atilde}).
By a change of summation and integration variables 
we can also symmetrise the expression with respect to 
$(R\leftrightarrow S, \beta\leftrightarrow\gamma)$, 
since $\tilde A_\beta, \tilde A_\gamma$ commute. Using 
in addition that 
\be
\sum_i g_i^3 f_0^{(i)} = 2g^3 N_f \,\delta_+(p^2)\, N_-(p_0), 
\ee
we arrive at 
\ba
j^\mu(Q) & = & g^3 N_f \int_{R,S}\int_{P} \delta_{Q-R-S} \,
\,\delta_+(p^2)\, N_-(p_0) \nn
& \times & 
\biggl[ 
Q^\alpha g^{\beta\gamma} + 
R^\beta g^{\gamma\alpha} \frac{p\cdot Q}{p\cdot R}+
S^\gamma g^{\alpha\beta} \frac{p\cdot Q}{p\cdot S}
\biggr] \frac{\partial}{\partial p^\alpha}
\biggl(
\frac{p^\mu}{p \cdot Q}\biggr) \tilde A_\beta (R) \tilde A_\gamma(S).
\la{eq:jmufin}
\ea

Let us now show that the same expression is obtained by computing
\be
j^\mu(Q) = \int_x e^{-i Q\cdot x} j^\mu(x) = 
- \int_x e^{-i Q\cdot x} \frac{\delta}{\delta A_\mu(x)} \delta S_M 
= -\int_{Q'}\delta_{Q+Q'} \frac{\delta}{\delta A_\mu(Q')} \delta S_M,
\ee
where we used 
${\delta A^\nu(Q')}/{\delta A_\mu(x)} = g^{\mu\nu} e^{-i Q'\cdot x}$, 
and we expect (cf.\ \eq\nr{s5})
\be
\delta S_M = - g^3 N_f \int_{Q,R,S} \int_{P}
\delta_{Q+R+S} \,\delta_+(p^2)\, N_-(p_0)  
\tilde A_\alpha(R) \tilde A^\alpha(S)
\frac{Q^\gamma}{p\cdot Q} \tilde A_\gamma(Q).  
\ee
Indeed, taking derivatives using the right-most side of \eq\nr{def} 
and writing the result again in terms of 
$\tilde A_\beta, \tilde A_\gamma$, 
we immediately recover~\eq\nr{eq:jmufin}. 
Thus the C odd operator in \eq\nr{s6} is 
reproduced by classical kinetic theory, at least in the Abelian case. 

It is interesting to note that to reproduce the action in~\eq\nr{s6}, 
we had to carry out two partial integrations in~\eq\nr{j3mu}. 
In contrast, the static limit, the last 
term in \eq\nr{abstatref}, is much more easily obtained
from the first row of~\eq\nr{j3mu}. We again postpone the 
discussion on this to the non-Abelian case in the next section. 

\subsection{Kinetic equations for the non-Abelian case}
\la{se:nonab}

To complete the discussion in the previous sections, we review
here briefly the form of the kinetic equations in the non-Abelian
case, and show that they reproduce the static limit discussed 
in~\se\ref{se:static}. Because of a considerable proliferation 
of formulae, without any additional physical insight involved
as far as we can judge, 
we do not here present a complete non-static analysis, though, 
as we did for the Abelian case. 

The simplest way to display the non-Abelian 
kinetic equations is to follow~\eq\nr{boltz}, 
but replace $f$ by an $N_c\times N_c$ matrix. The 
collisionless QCD Boltzmann equation 
(for finer details of terminology, see~\cite{uh,he})
for each single fundamentally charged fermionic degree of 
freedom, and the corresponding gauge current induced, 
can be written as
\ba
& & \Bigl[ p\cdot D, f\Bigr] + 
\frac{g}{2} \Bigl\{p^\mu F_{\mu\nu}, \frac{\partial f}{\partial p_\nu} \Bigr\}
 =  0 , \la{compact} \\
& & 
j_\mu^a = -g \int_{P} p_\mu \tr \Bigl[T^a f\Bigr].
\ea
To express this in a more familiar way, 
we may write the matrix $f$ in terms of 
a singlet distribution function $\bar f$ and an adjoint (or octet)
distribution function $f^a$:
\be
f(x,p) = \frac{1}{N_c} \bar f(x,p) + 2 T^a f^a (x,p).
\ee
Projecting then \eq\nr{compact} with $\tr [...]$
and $\tr[T^a ...]$, the equations obtain their usual forms
(with our sign conventions)~\cite{uh,he},
\ba
& &  
 p\cdot \partial \bar f  + g p^\mu F^a_{\mu\nu} 
 \frac{\partial f^a}{\partial p_\nu} = 0, \\
& & 
  (p\cdot {\cal D})^{ab} f^b 
+ \frac{g}{2} d^{abc} p^\mu F^b_{\mu\nu}  
  \frac{\partial f^c}{\partial p_\nu} 
+ \frac{g}{2N_c} p^\mu F^a_{\mu\nu} 
  \frac{\partial \bar{f}}{\partial p_\nu} = 0, \\
& & 
j_\mu^a  = -\sum_i g_i \int_{P} p_\mu f^{a(i)}. \la{jmua}
\ea
When computing the current, each quark flavour now comes 
with $N_c$ colours, in addition to two spin degrees of 
freedom, both for particles and for anti-particles.  

The equations can again be solved iteratively in $g F _ {\mu \nu}$: 
$f = f_0 + f_1 + f_2 + ...\,$. At the zeroth order, 
\ba
\bar f_0^{(i)} & = &  \delta_+(p^2)\, n_\rmi{F}(p_0 - \mu_i) ,  \\
f_0^{a(i)}  & = & 0.
\ea
Iterating this, we obtain at the first and second order,  
\ba
\bar f_1^{(i)} & = & 0 , \la{s1} \\
(p\cdot {\cal D})^{ab} f^{b(i)}_1 & = & 
-\frac{g_i}{2 N_c} p^\mu F^a_{\mu 0} \,\delta_+(p^2)\, 
n'_\rmi{F}(p_0-\mu_i), \la{a1} \\
p\cdot \partial \bar f_2^{(i)} & = & -g_i p^\mu F^a_{\mu\nu}
\frac{\partial f_1^{a(i)}}{\partial p_\nu}, \la{ss2} \\ 
(p\cdot {\cal D})^{ab} f^{b(i)}_2 & = & -\fr{g_i}{2} d^{abc} p^\mu F^b_{\mu\nu}
\frac{\partial f_1^{c(i)}}{\partial p_\nu}. \la{a2}
\ea

In order to argue that these equations contain the same physics 
as~\eq\nr{s6}, we shall complement 
the full proof for the Abelian case in~\se\ref{se:cubic}, 
by showing that the full non-Abelian static
limit of \eq\nr{statref} is also reproduced. 
To demonstrate the latter point, we start from 
the first order in $g F_{\mu\nu}$. In~\eq\nr{a1} we can write 
\be
p^\mu F^a_{\mu 0} = (p\cdot {\cal D} A_0)^a - \partial_0 (p^\mu A_\mu^a) = 
(p\cdot {\cal D})^{ab} A_0^b, 
\ee
so that 
\be
f_1^{a(i)} = -\frac{g_i}{2N_c} A_0^a\,\delta_+(p^2)\, 
n'_\rmi{F}(p_0-\mu_i).
\ee
Inserting into \eq\nr{jmua}, we obtain 
\ba
j_\mu^a = -\sum_i g_i \int_{P} p_\mu f_1^{a(i)} & = &  
g^2 N_f A_0^a \int_{P} p_\mu \,\delta_+(p^2)\, 
N'_+(p_0) \nn 
& = & \delta_{\mu 0}\, g^2 N_f A_0^a \int_\vec{p} 
N'_+(\omega_P).
\ea
After use of \eq\nr{int2}, this agrees with~\eq\nr{statref}.

The equation for $f^a_2$, on the other hand, becomes 
\ba
(p\cdot {\cal D})^{ab} f^{b(i)}_2 & = &  
\frac{g_i^2}{4N_c}d^{abc} p^\mu F^b_{\mu 0}  A_0^c \,\delta_+(p^2)\,
n''_\rmi{F}(p_0 - \mu_i) \nn
& = & 
\frac{g_i^2}{4N_c}d^{abc} (p\cdot {\cal D} A_0)^b  A_0^c \,\delta_+(p^2)\,
n''_\rmi{F}(p_0 - \mu_i).
\ea
Here
\be
d^{abc} (p\cdot {\cal D} A_0)^b  A_0^c = 
\fr12 (p\cdot {\cal D})^{ab} (d^{bcd} A_0^c A_0^d),
\ee
so that 
\ba
f^{a(i)}_2 & = & 
\frac{g_i^2}{8 N_c} d^{abc} A_0^b A_0^c \, 
\,\delta_+(p^2)\, n''_\rmi{F}(p_0 - \mu_i). 
\ea
Carrying out the sum over $i$ in \eq\nr{jmua}, we finally get
\ba
j_\mu^a & = & - \fr14 g^3 N_f d^{abc} A_0^b A_0^c 
\int_{P} p_\mu \,\delta_+(p^2)\, N_-''(p_0) \nn
 & = & - \delta_{\mu 0}\, \fr14 g^3 N_f d^{abc} A_0^b A_0^c
 \int_\vec{p} N_-''(\omega_P).
\ea
Using \eq\nr{int3}, this agrees with \eq\nr{statref}.

\subsection{Possible simplifications of the classical description?}

We have argued that classical kinetic theory should 
reproduce the results of   
the Hard Thermal Loop effective theory even for C odd
observables, thus correctly 
representing the infrared physics of the system. 
It has however some merits over the Hard Thermal Loop action: 
it is local, directly Minkowskian and, 
involving only the solution of classical
equations of motion, can be implemented numerically 
without the need for analytic continuation. 
Moreover, judging from non-perturbative
studies of dimensional reduction, such numerical results  
should be at least qualitatively reliable for $T\sim (2 T_c,\infty)$.

In full generality, the classical distribution functions
$f(x,p)$ depend on (4+4) coordinates. If one wants to proceed 
towards numerical implementation, it is best to reduce the 
dimensionality of the phase space. As we have discussed, 
$f(x,p) = \delta_+(p^2)\hat f(x,p)$, and the  
dependence can thus be reduced to, say, the spatial components 
$\vec{p}$. This (4+3) dimensional problem can already
be managed on the lattice~\cite{mhm}. However, for the usual C even 
case one can carry out a further simplification, by writing in~\eq\nr{a1} 
\be
f_1^{a(i)} = -\frac{g_i}{2 N_c} \,\delta_+(p^2)\,
n'_\rmi{F} (p_0 - \mu_i) \, W_1^a(x,\vec{v}), 
\ee
and performing
the sum over $i$ and the integral over $p_0$
in \eq\nr{jmua}, reducing thus the
dependence only to angular variables and a total of (4+2) dimensions.
Further simplifications in the angular variables (like an expansion in 
spherical harmonics~\cite{rh,bmr}) may also be possible. 

It is then natural to ask whether a similar simplification
could be made in the presence of $f_2^{a(i)}$. In order to have 
a proper statistical weighting of the initial conditions 
for the time evolution, one should also work out 
a Hamiltonian formulation in terms of the gauge fields 
$A_i^a$, $E_i^a \equiv F_{0i}^a$ and $W_1^a, W_2^a$, an issue which
is not altogether trivial~\cite{bi,nair}. We have not 
carried out these constructions, but are not
aware of any {\em a priori} fundamental problems in doing so.

\section{Conclusions}
\la{se:concl}

In this paper we have addressed the computation of real time
observables in QCD at finite temperatures and densities. 

Most of the physical observables one can think of are not 
computable in perturbation theory, due to infrared problems.
Real time observables are
not well suited for direct 4d lattice simulations, either. 
However, it is possible to use perturbation theory to construct 
an effective description of the infrared dynamics. If simple
enough, this could then be studied by 
non-perturbative means. 

We have here completed the bosonic sector of the Hard Thermal 
Loop effective theory for real time observables, by computing the 
leading C odd operator induced by a finite density. This operator
is of a relatively simple form (\eq\nr{s6})
and, as we have argued, 
is reproduced 
by classical kinetic theory, whose equations
of motion are local (\se\ref{se:nonab}).

The classical kinetic theory can then be studied on the lattice, 
as has been demonstrated in the case of the sphaleron rate
in the electroweak theory~\cite{mhm,bmr}, as well as in the case of
the defect formation rate in scalar electrodynamics~\cite{rh}. Strictly
speaking there are still problems in finding a formulation which 
has a well-defined continuum limit~\cite{db2}, 
but these may not be important for practical applications, 
in which one may hope to find an intermediate scaling window 
allowing to extract physical results.   

When the chemical potential is turned on, 
the further problem arises that the ``Hamiltonian'' 
determining the thermal distribution of 
initial conditions is complex, resulting in a ``sign problem'':
standard Monte Carlo methods are not efficient in sampling the 
configuration space, when one is close to the thermodynamic
(infinite volume) limit. Numerical tests in the static case
have shown, though, that in practice one can reach large 
enough volumes before the sign problem starts to reduce
the signal-to-noise ratio in any significant way~\cite{mu,reim}.

Thus, it seems that a numerical determination of many 
non-perturbative real time quark--gluon plasma observables may 
become feasible, using classical kinetic theory. 


\section*{Acknowledgements}
The work of M.L.\ was partly supported by the TMR network 
{\em Finite Temperature Phase Transitions in Particle Physics}, 
EU Contract No.\ FMRX-CT97-0122.
D.B.\ is supported by the U.S.\ Department of Energy, Contract No.\
DE-AC02-98CH10886.

\end{document}